\documentclass[seceq]{ptptex}

\usepackage{graphicx}

\title{
NA61/SHINE physics program -- first results and future plans%
}

\author{
Tobiasz \textsc{Czopowicz} (for the NA61/SHINE Collaboration)%
}

\inst{
Faculty of Physics, Warsaw University of Technology, Poland
}

\abst{
The NA61/SHINE experiment aims to discover the critical point of strongly interacting matter and study properties of the onset of deconfinement.
These goals are to be achieved by performing a two-dimensional phase diagram ($T-\mu_{B}$) scan -- measurements of hadron production as a function of collision energy and system size.
With its large acceptance and good particle identification NA61/SHINE also performs detailed and precise particle production measurements for the T2K, Pierre Auger Observatory and KASCADE-Grande experiments.
This contribution summarizes current status and future plans as well as presents the first physics results of the NA61/SHINE experiment.
}

\begin{document}

\maketitle

\section{The NA61/SHINE physics goals}
NA61/SHINE ({\bf S}PS {\bf H}eavy {\bf I}on and {\bf N}eutrino {\bf E}xperiment) \cite{proposal} is a fixed-target experiment located in the North Area of the Super Proton Synchrotron (SPS) accelerator facility at the European Organization for Nuclear Research (CERN) in Geneva, Switzerland.
It is the successor of the NA49 experiment \cite{Afanasiev:1999iu}, which was operating in 1994 -- 2002.
The NA61/SHINE collaboration consists of 140 physicists from 28 institutes in 13 countries.
NA61/SHINE physics goals are:
\begin{itemize}
	\item search for the critical point of strongly interacting matter,
	\item detailed study of the onset of deconfinement,
	\item study high $p_T$ hadrons in p+p and p+Pb interactions,
	\item perform reference measurements for neutrino physics (the T2K experiment) by measuring the hadron production of the T2K target exposed to the proton beam at 31$A$~GeV/c\cite{Strabel:2010qp, Palczewski:2009pq},
	\item perform reference measurements for cosmic-ray physics (KASCADE-Grande and Pierre Auger Observatory experiments) by measuring p+C and $\pi$+C interactions \cite{arXiv:1012.2604, Abgrall:2010fm}.
\end{itemize}

\begin{figure}[h]
	\centering
	\includegraphics[width=0.5\textwidth]{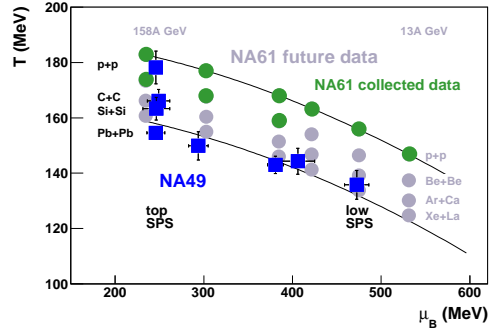}
	\caption{Freeze-out points of the planned NA61/SHINE two-dimensional phase diagram scan (based on \cite{Becattini:2005xt}). Be+Be at three highest energies have been recorded in November 2011, already after the conference.}
	\label{fig:NA61_phase_diagram_scan}
\end{figure}

In order to search for the critical point and study the properties of the onset of deconfinement, NA61/SHINE performs a two-dimensional phase diagram scan.
It measures hadron production in various collisions (p+p, Be+Be, Ar+Ca, Xe+La) at various beam energies (13$A$, 20$A$, 30$A$, 40$A$, 80$A$ and 158$A$~GeV) \cite{add-5}.
These new data, together with Pb+Pb reactions recorded by NA49 will allow to cover the region, where the critical point is expected (Fig.~\ref{fig:NA61_phase_diagram_scan}).
According to lattice calculations $T^{CP} = 162 \pm 2$ MeV and $\mu^{CP}_{B} = 360 \pm 40$ MeV \cite{Fodor:2004nz}.
NA61/SHINE will search for the onset of ``kink, horn, step'' \cite{Gazdzicki:1998vd, arXiv:0710.0118} signatures in light nuclei and a maximum of fluctuation signals for systems freezing-out close to the critical point.

\section{Detector overview}
\begin{figure}[h]
	\centering
	\includegraphics[width=0.75\textwidth]{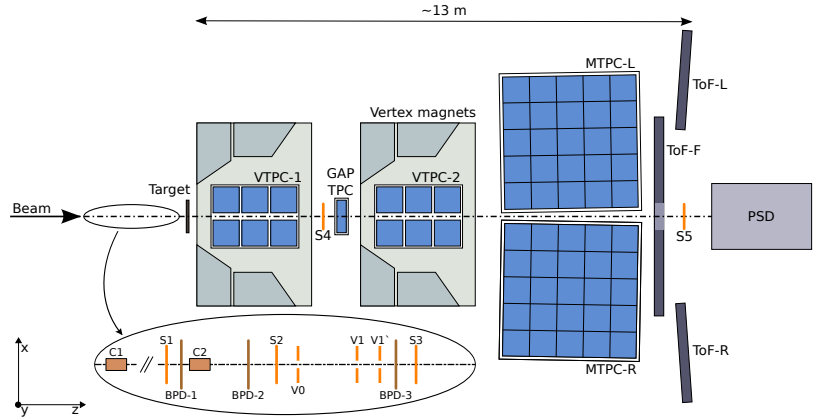}
	\caption{The layout of the NA61/SHINE experimental set-up (top view, not to scale). For details see\cite{Afanasiev:1999iu}.}
	\label{fig:NA61_setup}
\end{figure}

The NA61/SHINE detector is a large acceptance hadron spectrometer.
The main components of the current detector were constructed and successfully used by NA49.
The experimental set-up is shown in Fig.~\ref{fig:NA61_setup}.

The main tracking devices are four large volume Time Projection Chambers (TPC) inherited from NA49.
Two of them are located in the magnetic field, two others are positioned downstream of the magnets symmetrically to the beam line.
These detectors allow reconstruction of over 1000 tracks in a single Pb+Pb interaction.

The most downstream detector (the Projectile Spectator Detector, mounted in 2011) is designed to measure the number of non-interacting nucleons from the projectile nuclei on event-by-event basis.
High resolution of measurement of the total energy of the projectile spectators ($\sigma(E)/E < 55\%/\sqrt{E}$) in a very broad energy range should lead to a low uncertainty in the determination of the number of interacting nucleons even for peripheral collisions of heavy nuclei at low energies.

\section{Data taking}
During the first two years of running over 120 million events have been collected (including data for neutrino and cosmic-ray physics).
Also the first steps of the phase diagram scan were accomplished: p+p data at six energies were collected (Fig.~\ref{fig:NA61_phase_diagram_scan}).
In 2011 first ion run -- partial energy scan with Be+Be collisions -- took place.
The program will be completed by Ar+Ca and Xe+La energy scans in 2014 and 2015, respectively.
The remaining three energies of Be+Be will be taken in 2012.

\section{First results}
\begin{figure}[h]
	\centering
	\includegraphics[width=0.75\textwidth]{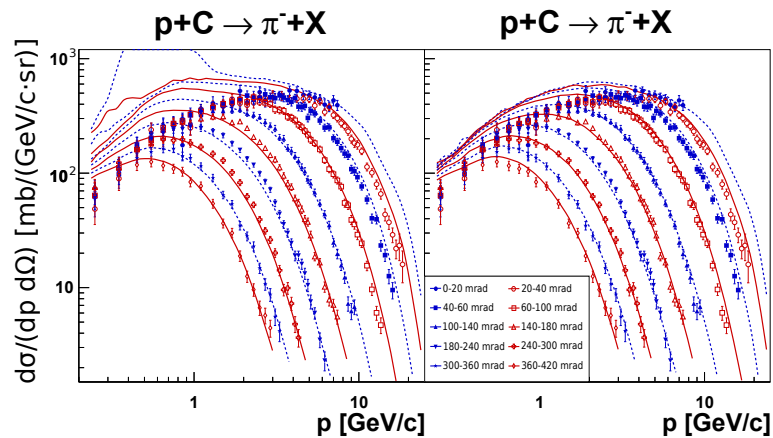}
	\caption{Inclusive charged pion spectra measured in NA61/SHINE (points) compared to UrQMD 1.3.1 (lines) (left) and patched UrQMD 1.3.1 (right).}
	\label{fig:NA61-URQMD}
\end{figure}

The first precise measurements of charged $\pi$ spectra in p+C interactions at 31$A$ GeV recorded in 2007 have been published\cite{arXiv:1102.0983}.
These results have already been used by the T2K experiment for the calculations of the neutrino flux\cite{arXiv:1106.2822}, which allowed to measure $\theta_{13}$ parameter of neutrino oscillation matrix (MNS).
Also, comparison with the UrQMD model revealed discrepancies at low momenta and low azimuthal angles (Fig.~\ref{fig:NA61-URQMD}).
Taking into account the NA61/SHINE results, a patch has been released to correct the model\cite{arXiv:1109.6768}.

The measurements of $K^{+}$ spectra from p+C interactions at 31$A$~GeV have just been released\cite{arXiv:1112.0150}.

\begin{figure}[h]
	\centering
	\includegraphics[width=0.45\textwidth]{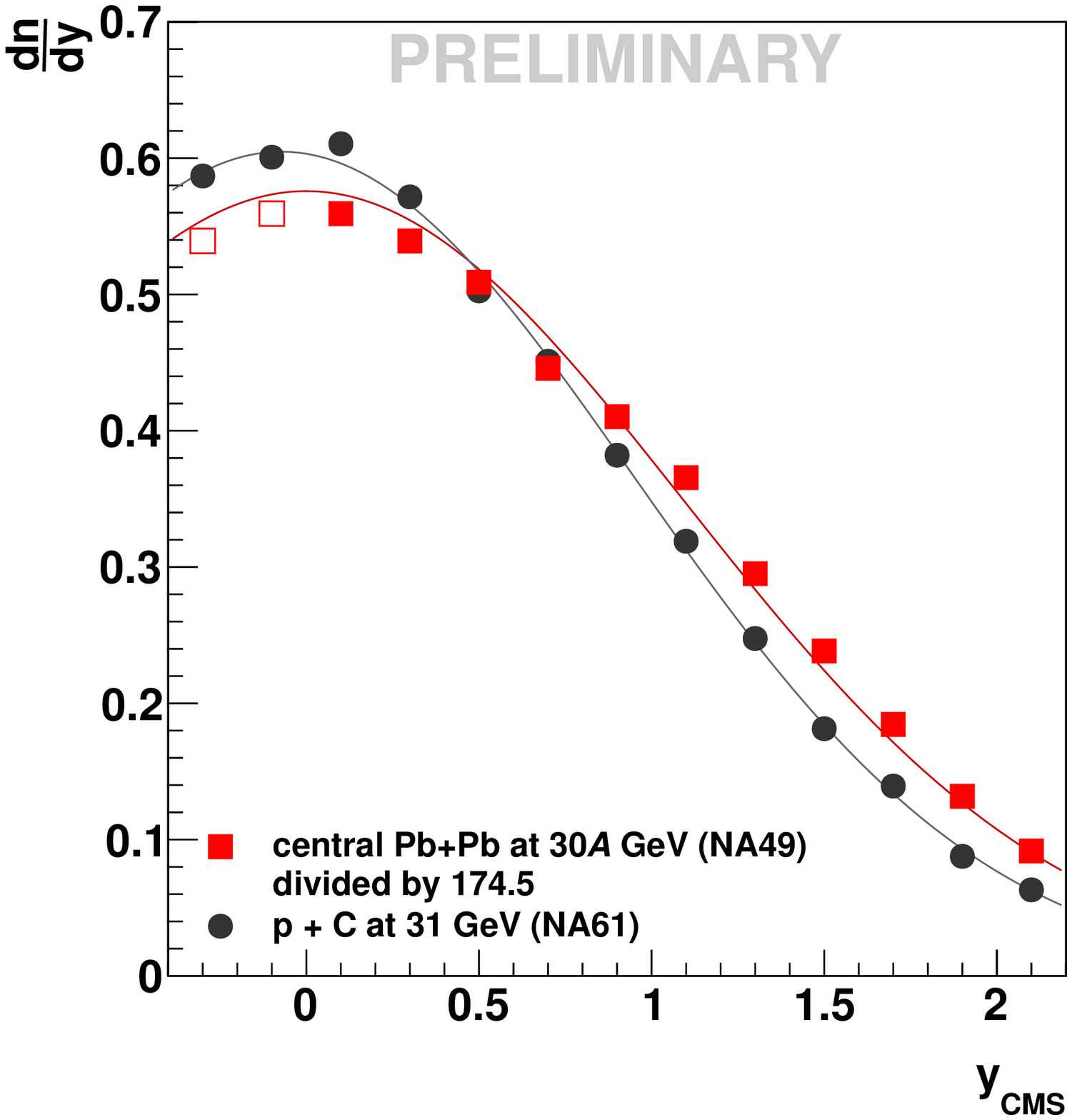}
	\includegraphics[width=0.45\textwidth]{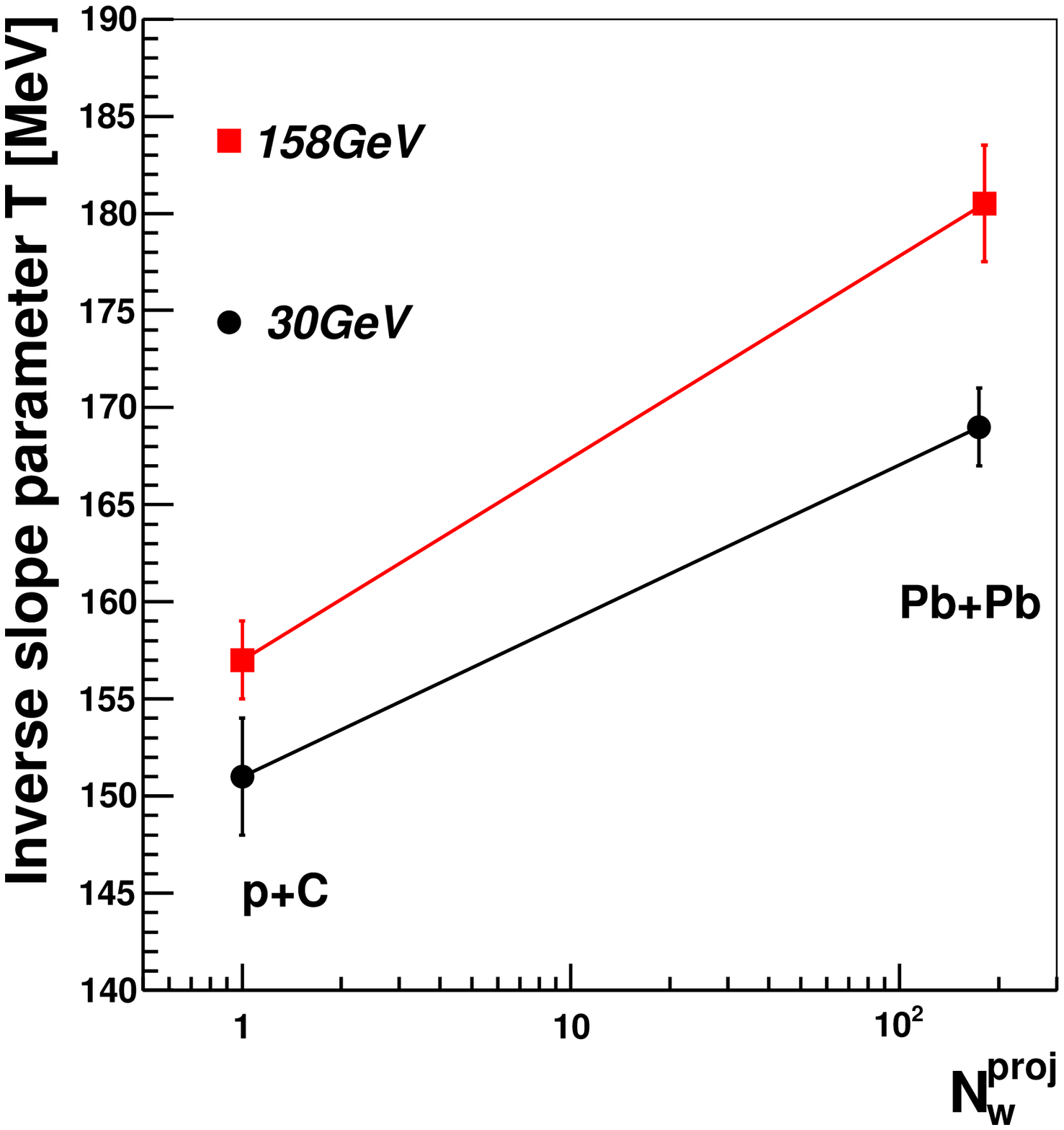}
	\caption{$\pi^{-}$ measurements: (left): rapidity distribution in p+C at 31$A$~GeV and central Pb+Pb interactions at 30$A$~GeV for $p_{T} < 1~GeV/c$.
			The Pb+Pb points are scaled by the predicted wounded projectile nucleon ratio in these two reactions.
			(Right): inverse slope parameter of the transverse mass distribution at mid-rapidity for p+C and Pb+Pb at 30$A$/31$A$ and 158$A$~GeV.}
	\label{fig:NA61_results}
\end{figure}

First $\pi^{-}$ results relevant to the study of the properties of the onset of deconfinement are presented in Fig.~\ref{fig:NA61_results}.
The rapidity spectra from p+C interactions at 31$A$~GeV and central Pb+Pb collisions at 30$A$~GeV are compared in Fig.~\ref{fig:NA61_results} (left).
The mean pion multiplicity in the forward hemisphere is approximately proportional to the mean number of wounded nucleons of the projectile nucleus.
At the energy of the onset of deconfinement (30$A$~GeV) the mean pion multiplicity in central Pb+Pb collisions agrees with the Wounded Nucleon Model\cite{TPJU-9/76} predictions.

An exponential function was fitted to the transverse mass spectra at mid-rapidity ($0 < y < 0.2$) in the range of $0.2 < m_{T} - m_{\pi} < 0.7~GeV/c^{2}$.
The fitted inverse slope parameter T is plotted in Fig.~\ref{fig:NA61_results} (right) for p+C at 31$A$~GeV (this analisis), 158$A$~GeV\cite{hep-ex/0606028}, and central Pb+Pb at 30$A$\cite{arXiv:0710.0118} and 158$A$~GeV\cite{arXiv:0710.0118}.
An increase is observed both when increasing the system size, and when the collision energy.
In statistical models this behaviour can be interpreted as a temperature and radial flow increase.

\section{Summary}
NA61/SHINE has the potential to discover the critical point of strongly interacting matter and guarantees systematic data on the onset of deconfinement.
The two-dimensional scan of the phase diagram is in progress.
p+p interactions were already registered at all six energies.
First Be+Be run took place in November 2011.
Charged pion spectra have been used for computation of the T2K $\nu_{\mu}$ beam properties, as well as for improving the UrQMD model.


%

\end{document}